# Exploring Perceptions of Veganism

Laura Jennings[a], Christopher M. Danforth[b], Peter Sheridan Dodds[b], Elizabeth Pinel[c], Lizzy Pope[a*]

**Abstract:** This project examined perceptions of the vegan lifestyle using surveys and social media to explore barriers to choosing veganism. A survey of 510 individuals indicated that non-vegans did not believe veganism was as healthy or difficult as vegans. In a second analysis, Instagram posts using #vegan suggest content is aimed primarily at the female vegan community. Finally, sentiment analysis of roughly 5 million Twitter posts mentioning "vegan" found veganism to be portrayed in a more positive light compared to other topics. Results suggest non-vegan's lack of interest in veganism is driven by non-belief in the health benefits of the diet.

[a] Department of Nutrition and Food Sciences, College of Agriculture and Life Sciences, University of Vermont
[b] Department of Mathematics and Statistics, College of Engineering and Mathematical Sciences, Vermont Complex Systems Center, University of Vermont
[c] Department of Psychological Sciences, College of Arts and Sciences, University of Vermont

Corresponding Author:

Lizzy Pope, PhD RDN
Dept. of Nutrition and Food Sciences
University of Vermont
254 Carrigan Wing
109 Carrigan Drive
Burlington, VT 05405
802-656-4262
802-656-0001
efpope@uvm.edu

**Introduction**

Veganism is a style of eating where followers abstain from eating any animal products. Research supports multiple health benefits of following a vegan diet (Johansson & Larsson, 2002). Because the vegan diet excludes all animal products, individuals whom are vegan typically consume a greater amount of whole, plant-based foods (Johansson & Larsson, 2002). In fact, a recent study determined that total vegetable, legume, and dietary supplement consumption is higher in vegans compared to omnivores (Johansson & Larsson, 2002). Subsequently, previous research indicates that vegans have a lower energy intake compared to individuals following other diets (Clarys et al., 2014). Furthermore, compared to omnivorous, vegetarian, and semi-vegetarian diets, the prevalence of being overweight or obese is lowest for vegans (Clarys, et al., 2014). This association may exist because in addition to higher quality diets, vegans also often practice healthful behaviors such as regular exercise, not smoking, and minimal alcohol consumption (Dyett, Sabaté, Haddad, Rajaram, & Shavlik, 2013).

The improved diet quality and healthful behaviors associated with veganism may also contribute to several morbidity benefits. Cardiovascular disease, the nation's leading cause of death, is less prevalent in vegans. A recent study found that vegan males have a 55% reduced risk for developing heart disease (Le & Sabaté, 2014). Furthermore, vegans have a 75% reduced risk of developing hypertension, which can lead to cardiovascular disease (Le & Sabaté, 2014). Additionally, one study found that 93% of cardiovascular disease patients experienced improvement or resolution of symptoms after following a vegan diet (Esselstyn Jr, Gendy, Doyle, Golubic, & Roizen, 2014). This suggests that in addition to the prevention of chronic diseases, this diet also has reversal effects. After heart disease, cancer is the second leading cause of death in the United States (Prevention, 2017). A recent meta-analysis determined that adherence to a vegan diet significantly reduced the risk of total cancers (Dinu, Abbate, Gensini, Casini, & Sofi, 2017). Type 2 diabetes is another health concern with increasing prevalence. However, vegans have been found to have a 47-78% reduced risk of developing type 2 diabetes compared to non-vegetarians (Le & Sabaté, 2014). Taken together, these studies indicate that following a vegan diet can have a positive impact on one's health.

Despite the positive health benefits associated with following a vegan diet, very few Americans have chosen to adopt veganism as a way of eating (Johansson & Larsson, 2002). According to a survey conducted in 2014, only 0.5% of the population is considered to be a current vegan (Asher et al., 2014). Another recent survey of former vegetarians/vegans suggests that the low-rate of vegan adherents could be due to the inconvenience of being vegan and the restricted food options (Asher, et al., 2014). A second obstacle noted in the survey was the social aspect of the diet and how it can make some feel isolated in social settings (Asher, et al., 2014). This isolation may partly account for the large presence of veganism on social media, as it may provide vegan users a sense of community. Social media is a growing venue for food-related posting, and a place where people express opinions on many topics. The use of all forms of social media (Instagram, Twitter, Snapchat, Facebook, etc.) is increasing each year not just in frequency of use, but also in number of users (Smith & Anderson, 2018). However, there is skepticism about the extent to which behavior portrayed on social media is an accurate representation of societal beliefs. One study found that oftentimes, public opinion evolves into an ordered state so that one opinion seems to dominate all others on a certain platform (Xiong & Liu, 2014). Other work has found strong correlations between public opinion and Twitter sentiment, including for example Presidential approval ratings. During his two terms in office, President Obama's quarterly approval rating was strongly correlated with the sentiment of tweets mentioning his name during the preceding 3 months (Cody, Reagan, Dodds, & Danforth, 2016). The aim of this study is to determine the perception of veganism portrayed on social media, and how this may differ or resemble what peoples' perceptions of veganism are outside of social media. Gaining a better understanding of perceptions of veganism may help public health officials encourage more Americans to at least approximate a vegan diet.

**Methods**

*Vegan Perception Survey*

The first part of this study involved a survey that was completed by a mix of 500 vegans and non-vegans using Amazon's Mechanical Turk (MTurk). MTurk is a crowdsourcing web service that can be used to gain access to a diverse population of survey participants. Researchers can use MTurk to gather data from workers who are then compensated in return (Amazon, 2018). The use of sites such as these on the internet has been shown to be an effective means of collecting data regarding perceptions than in person sampling. Not only is it a more efficient way to collect a large amount of data, but it also allows for access to a more representative sample of participants compared to lab-based research (Woods, Velasco, Levitan, Wan, & Spence, 2015). The survey contained three different versions, and a participant's experience with veganism determined which version of the survey they completed. One version was for those that were vegan, one was for those that were non-vegans that were not considering veganism, and one was for non-vegans that were considering veganism. Each version addressed factors such as health habits, food consumption, and lifestyle of the participants. Additionally, beliefs of what being a vegan entails were examined. Some questions were asked in all three versions and were answered on a Likert scale from 1-10 for easy comparison. For example, "How difficult do you think it is to maintain a vegan diet?" and "How healthy do you think the vegan diet is compared to a diet that includes animal products?" were asked in all versions. Not only did this allow for the comparison of different lifestyles, but it also showed how non-vegans perceive vegans and vice versa. This provided a baseline as to what the perception of veganism is in the public by both vegans and non-vegans. A copy of the survey can be found in Appendix 1. The study protocol was approved by the Committee on Human Research in the Behavioral and Social Sciences at the University of Vermont.

Survey data was analyzed using SPSS version 25 (IBM, Chicago, IL). Descriptive analyses were performed for all survey variables. To compare differences between vegans, those considering veganism, and non-vegans on survey variables, one-way ANOVA analyses were run. Qualitative data from the survey was compiled into one document, and used to support themes illuminated by the quantitative data.

*Instagram*

The method that was used to gather data from Instagram was based on a previous study by Tiggeman et al., on #fitspiration, and involved obtaining a sample of 600 Instagram posts associated with the "vegan" hashtag (Tiggemann & Zaccardo, 2018). These posts were then hand-coded into categories according to what they portrayed. Three coders worked together to code all images. Coders came to a consensus on each image before moving onto the next image. Initially, images were coded by content category: people, food, cosmetics, quotes, promotional, other. Images featuring food or people were then coded further. First, those featuring food were coded into one of four categories: everyday foods, everyday dessert foods, indulgent food porn, or healthy food porn. Everyday foods described posts that were of average photo quality of typical meals. Everyday dessert foods described posts which were of average photo quality of desserts. Indulgent food porn described posts of high photo quality of food with low nutritional value. Finally, healthy food porn described posts of high photo quality of food with high nutritional value. Posts featuring people were coded by number of people, gender, body type, and activity. First, the image was classified as showing one person or more than one person. The gender of the people in the photo was also coded. Secondly, the body shape of the person in the photo was coded as either fit, average, larger bodied, or unknown. Thirdly, the activity the person was doing in the photo was coded. A more thorough presentation of coding categories can be found in Table 1.

**Table 1.** Description of Coded Variables on Instagram

| Variable | Description | Details |
|---|---|---|
| Category | People, food, other, or irrelevant | • Images of people featured one or more person(s)<br>• Images of food included a food item<br>• All other images related to veganism were coded as 'other', while those unrelated to veganism were coded as irrelevant |
| Food | The quality and type of food | • Indulgent food porn - high quality photo of food with low nutritional value, e.g., ice cream<br>• Healthy food porn - high quality photo of food with high nutritional value, e.g., smoothie<br>• Everyday foods - average quality photo of typical meals, e.g., oatmeal<br>• Everyday dessert food - average quality photo of desserts, e.g., cake |
| Gender | The gender of the individual(s) | • Male<br>• Female<br>• Both - where two or more people of different genders are present |
| Body Type | The individual's physical build | • Fit - Little to no visible fat stores, visible muscle definition<br>• Average - moderate level of visible fat and muscle<br>• Larger Body - high level of visible fat, no visible muscle definition<br>• Unknown - unable to determine due to framing of image or clothing covering body |
| Activity | The activity the individual is carrying out in the image | • Selfie - image taken by individual in the frame<br>• Being active - engaging in a fitness activity<br>• Eating - engaging in eating food<br>• Fitspiration - modeling that is fitness related<br>• Food pose - posing with food<br>• Modeling - modeling that is fitness unrelated |
| Other | Vegan related posts which do | • Quotes - post containing text<br>• Promotional - advertisements for products or companies |

|  | not fit into previously defined categories | • Animals - post containing animals<br>• Cosmetics - post related to cosmetic products<br>• Education - post providing educational material<br>• Cookbooks - post features vegan cookbooks |
|---|---|---|
| Irrelevant | Does not fit into any categories | • The post is unrelated to veganism |

*Twitter*

Twitter data was collected through the Decahose streaming API, a feed providing a random 10% of all public messages. For the present study, the stream was subsampled to focus on messages mentioning the word "vegan". Using sentiment analysis, the average happiness of these tweets was calculated for each day from September 9, 2008 through November 2015. The sentiment was calculated using the Hedonometer algorithm. The tool looks at the most frequently used 10,000 words in the English language, referred to as labMT words, which have been given a happiness score from 1-9 to allow for comparisons (Cody, et al., 2016). Amazon's Mechanical Turk was used to score 5,000 of the most frequent words from Google Books, New York Times articles, Music Lyrics, and Twitter messages, which resulted in a composite set of roughly 10,000 unique words (Cody, et al., 2016). For example, the word "love" was rated by 50 participants to have a higher happiness score (8.42) than the word "hate" (2.34). Tweets mentioning the word "vegan" were scored based on the average happiness score of the words written in the tweet. 23,822,293 labMT words were found to co-occur in tweets mentioning "vegan". Assuming an average of five labMT words per tweet, this implies that there were roughly 5 million tweets in the sample. Using the Hedonometer, we determined which days were "happier" for veganism as well as which words accounted for the change. Additionally, these scores were compared against the average happiness scores of all posts that day, including those not mentioning "vegan." Ultimately, this determined whether or not veganism is portrayed in a more positive light on Twitter compared to other posts on Twitter that do not use the word vegan.

**Results**

*Survey*

Five-hundred and ten participants completed surveys using MTurk. Sixty-one percent of the sample was male, 35% was female, 2% was transgender male, 1% was transgender female, and 1% was non-binary. There were no significant gender differences between the vegan, considering veganism, or non-vegan groups, $\chi^2(8)=4.9$, $p=0.07$. Sixty-two percent of the sample was white, 9% was black, 26% was Asian, 4% was American Indian, Native American or Alaskan Native, 2% was Middle Eastern, 2% was Native Hawaiian or Other Pacific Islander, 1% was "other," and 4% was Hispanic or Latino. Participants were allowed to designate more than one race/ethnicity category. One-hundred and eighty-seven participants reported being vegan, 76 were sometimes vegan, and 247 were not vegan. The average age of participants was 33.3 with a range from 18 years old to 77 years old and a standard deviation of 9.66 years. Those who reported sometimes being vegan were directed to the version of the survey for vegans. Altogether, 263 participants completed the survey for vegans, 107 participants completed the survey for those considering veganism, and 140 participants completed the survey for non-vegans. The average length of veganism reported by vegans was 7.72 years, $SD=8.74$, N=253. Of the vegan participants, 73.4% reported health as the primary reason they decided to go vegan, 17.5% reported animal rights, 4.9% reported the environment, 2.3% reported outside influence, and 1.0% reported another primary reason.

There were significant differences between the survey groups on all variables of interest, see Table 2. Those considering veganism thought that the vegan diet was significantly more difficult than vegans or non-vegans, while vegans felt it was more difficult than non-vegans did. In terms of healthiness and expense vegans and those considering veganism believed the vegan diet was significantly healthier and more expensive than non-vegans. Vegans thought that media coverage was more positive than those considering veganism did or non-vegans. Similarly, vegans also felt that media coverage was driving more people to become vegan than those considering veganism or non-vegans did. Vegans thought that vegans were more offended by non-vegans than those considering veganism or non-vegans did, and vegans also thought that vegans were more offensive to non-vegans than those considering veganism or non-vegans believed. Vegans and those considering veganism rated vegan food as significantly healthier than non-vegans rated it. Non-vegans rated vegans as significantly less masculine than those considering veganism or vegans.

**Table 2.** Mean and SD for Dependent Variables by Vegan Status (Complete Dataset N=510)

| Social Variable | Vegans N= 263[a] | Considering Veganism N= 107 | Non-Vegans N= 140[b] | F (df, df) | p | Pairwise Test of Significance[c] |
|---|---|---|---|---|---|---|
| Vegan Diet Difficulty | 6.9 (2.3) | 8.0 (1.8) | 6.3 (2.2) | 19.37 (2, 502) | <0.001 | All comparisons significant |
| Vegan Healthy Diet | 7.6 (1.9) | 7.2 (1.8) | 4.0 (1.7) | 189.2 (2, 506) | <0.001 | V>N; C>N |
| Vegan Diet Expensive | 7.1 (2.1) | 6.6 (2.2) | 4.6 (2.0) | 61.3 (2, 504) | <0.001 | V>N; C>N |
| Vegan Media Coverage Tone | 6.9 (1.9) | 5.6 (1.9) | 3.9 (1.6) | 126.3 (2, 503) | <0.001 | All comparisons significant |
| Vegan Media Coverage Influence | 6.8 (2.2) | 4.8 (2.3) | 3.0 (1.5) | 157.5 (2, 505) | <0.001 | All comparisons significant |
| Vegans Offended by Non-Vegans | 6.7 (2.2) | 5.7 (2.2) | 4.1 (1.9) | 67.9 (2, 507) | <0.001 | All comparisons significant |
| Non-Vegans Offended by Vegans | 6.5 (2.3) | 5.5 (2.4) | 3.0 (1.7) | 121.0 (2, 507) | <0.001 | All comparisons significant |
| Vegan Food Healthier | 7.6 (1.9) | 7.2 (1.9) | 3.9 (1.8) | 189.1 (2, 505) | <0.001 | V>N; C>N |
| Vegans Less Masculine | 6.7 (2.2) | 5.9 (2.3) | 3.9 (2.0) | 77.5 (2, 505) | <0.001 | All comparisons significant |

[a] Number of vegan respondents per question ranged from 259-263
[b] Number of non-vegan respondents per question ranged from 138-140
[d] Pairwise significance tests via Bonferroni post-hoc, $p<0.05$

*Instagram*

Most of the posts fell into the food category (42.5%), followed by other (38.7%), people (11.8%), and irrelevant (7%). The images categorized as 'people' contained mostly female subjects (62.0%), and the fit body type was most prevalent (54.9%), with selfies being the most for various products or companies (60.3%), followed by cosmetic related posts (15.1%) and common activity (31%). See Table 3 for detail. Of the food images, most (58.0%) were of everyday foods, such as fruits and stir fries. The distribution of the other food categories can be seen in Table 3. The images which were categorized as 'other' were mostly promotional posts pictures of animals (10.8%). The distribution of the remaining subcategories can be seen in Table 3.

**Table 3.** Distribution of Subcategories of Instagram Posts

| Category | Subcategory | Total (%) |
|---|---|---|
| **Gender** | | |
| | Male | 16 (22.5) |
| | Female | 44 (62.0) |
| | Both | 11 (15.5) |
| **Body Type** | | |
| | Fit | 39 (54.93) |
| | Average | 19 (26.76) |
| | Inactive | 7 (9.86) |
| | Unknown | 6 (8.45) |
| **Activity** | | |
| | Selfie | 22 (31.0) |
| | Being Active | 15 (21.1) |
| | Eating | 1 (1.4) |
| | Food Pose | 11 (15.5) |
| | Fitspiration | 4 (5.6) |
| | Modeling | 18 (25.4) |
| **Food Description** | | |
| | Indulgent Food Porn | 32 (12.5) |
| | Healthy Food Porn | 55 (21.6) |
| | Everyday Foods | 148 (58.0) |
| | Everyday Dessert Foods | 20 (7.8) |
| **Other** | | |
| | Quotes | 22 (9.5) |
| | Promotional | 140 (60.3) |
| | Animals | 25 (10.8) |
| | Cosmetics | 35 (15.1) |
| | Education | 9 (3.9) |
| | Cookbooks | 1 (0.4) |

*Twitter*

      The daily happiness measurements of tweets containing the word "vegan" were found to be, on average, substantially higher than the daily happiness of all tweets. According to Figure 1, the average daily happiness of tweets containing the word "vegan" appears to oscillate around a happiness score of about 6.3. As a reference point, only a handful of days in the last decade have reached this level for all tweets (see Figure 2). The average daily happiness of tweets overall oscillates around a happiness score of 6.0. The day with the highest happiness rating for tweets containing "vegan" was 4/7/2015, with a score of 6.93. According to Figure 3, the words "love," "happy," and "birthday" were used more often on this day, contributing to this high score. Ariana Grande's tweet "happy birthday to my friend / gorgeous, honeymoon violin playing, vegan dream babe @TheKiaraAna ! i love u!" from this day is likely responsible for the large uptick, as it was retweeted 9,979 times. The day with the lowest happiness rating for tweets containing "vegan" was 10/27/2012, with a score of 5.24. According to Figure 4, the words "hate," "no," and "wait" were used more often on this day, contributing to this lower score. "I hate hipsters. Their smug faces, vegan diet, tiny feet & sawdust bedding. No wait. Hamsters. I hate hamsters." was retweeted 7,802 times, which could explain the lower score.

**Figure 1.** Daily Vegan Happiness

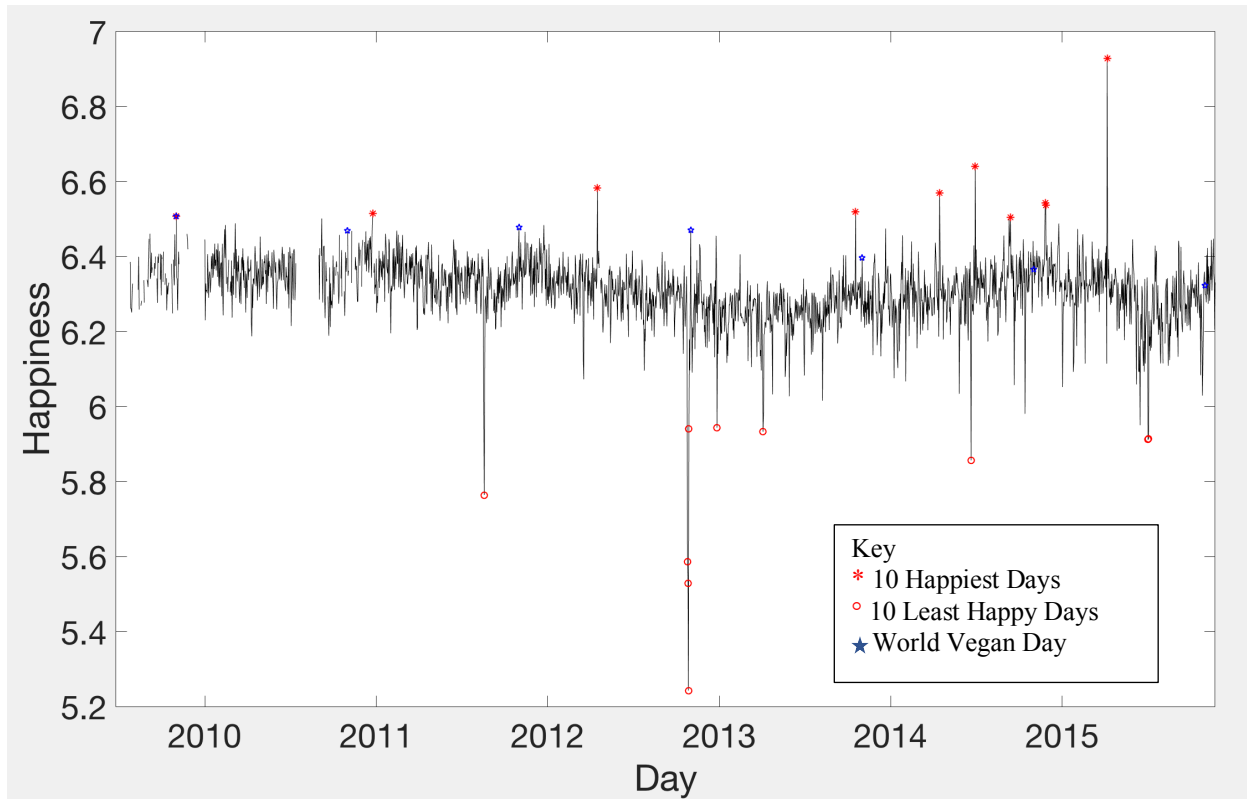

**Figure 2**. Average Happiness (Cody, et al., 2016)

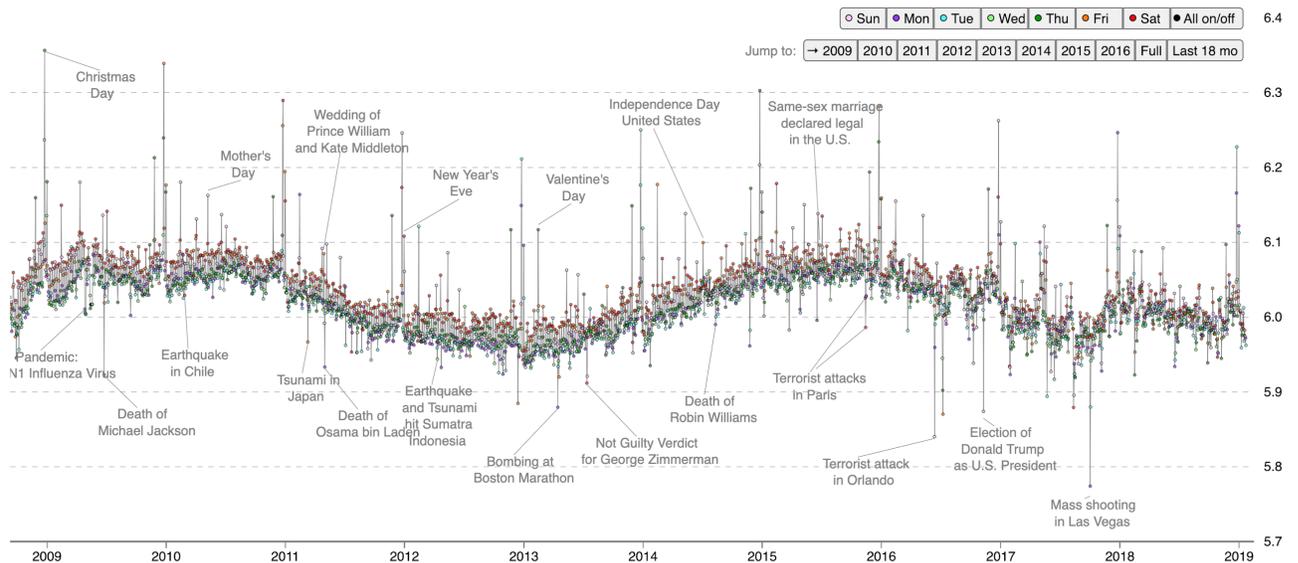

**Figure 3.** Word Shift of Happiest Vegan Day. This figure reveals individual word contributions to the difference in sentiment for 4/7/2015 (T_comp) as compared to the prior week (T_ref). Words appearing on the right contribute positively to this difference. For example, the relatively positive words "love", "happy", and "gorgeous" appear more often, and the relatively negative words "no", "not", and "don't" appear less often. Words appearing on the left go against the trend, with the relatively positive words "free", "food", and "best" appearing less often on 4/7/2015, and the relatively negative words "my" and "bad" appearing more often that day. Ambient Words for 'Vegan' on 4/7/2015; Average Happiness = 6.93

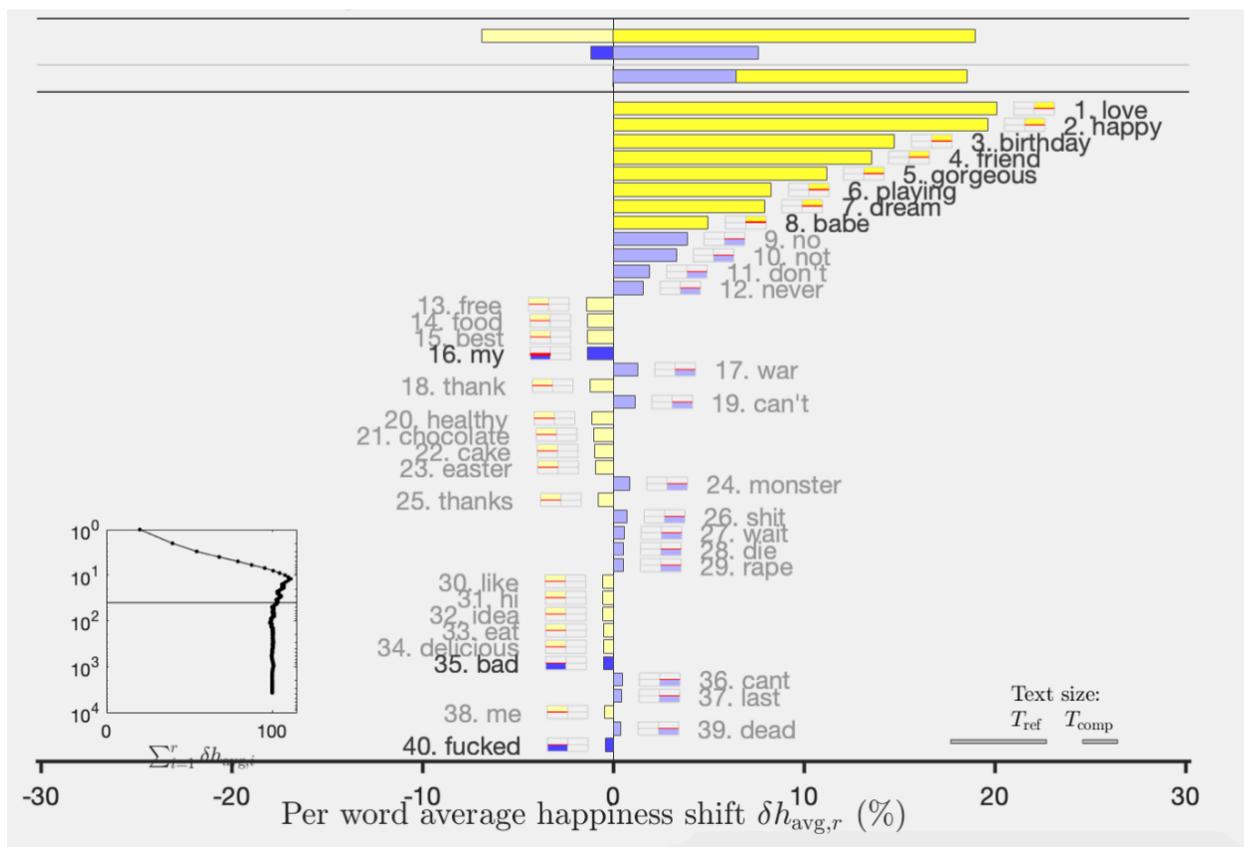

**Figure 4.** Word Shift of Least Happy Vegan Day

Ambient Words for 'Vegan' on 10/27/2012; Average Happiness = 5.24

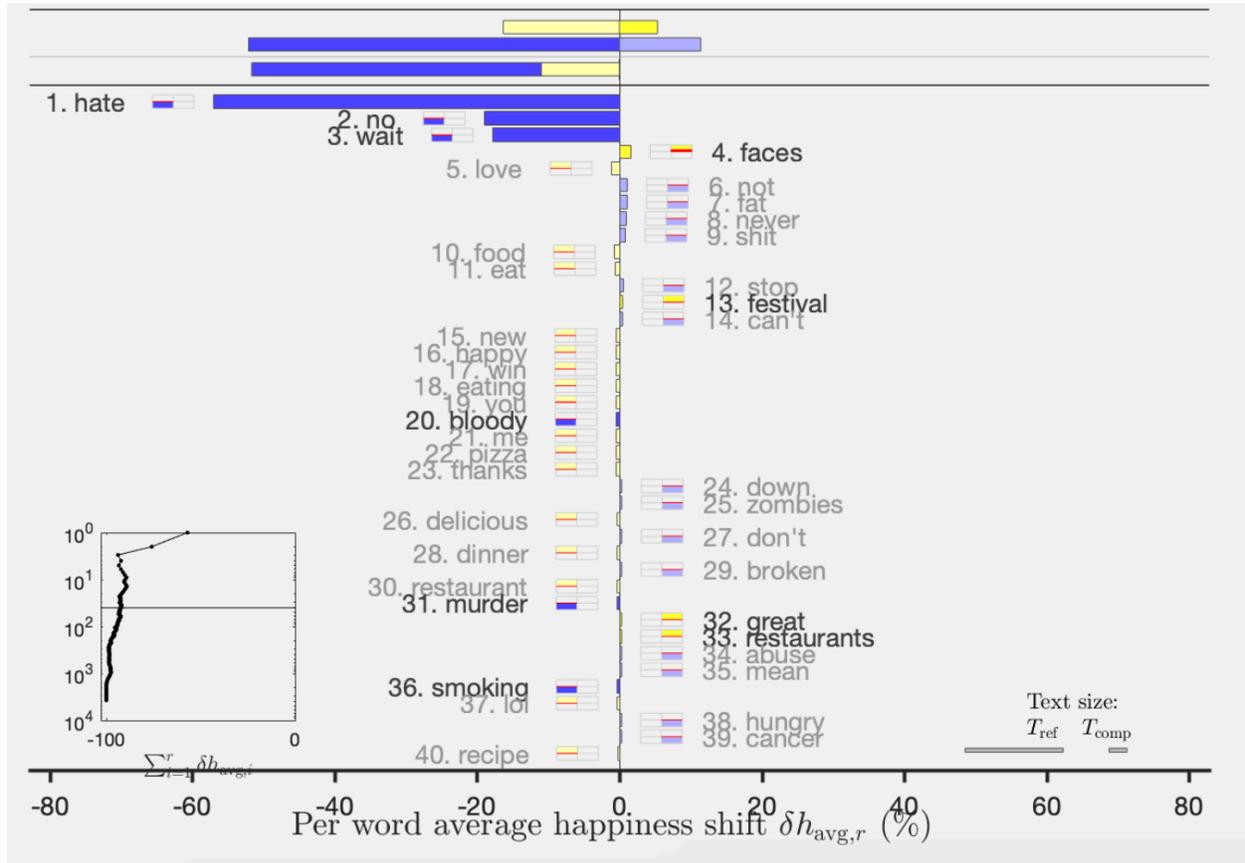

**Discussion**

    This study aimed to gather an overarching view of veganism in several formats to determine possible explanations as to why so few people have decided to be vegan. From the survey results, it was determined that health seems to be the primary reason that vegans decide to be vegan. However, non-vegans rated the vegan diet as less healthy compared to vegans and those considering veganism, which could explain why non-vegans lack motivation to become vegan. If health is the primary driver for most vegans and non-vegans do not even consider the diet to be

healthy, this could be one explanation why so many are choosing not to be vegan. This would indicate that there is a lack of education regarding the vegan diet for non-vegans.

A possible explanation for why non-vegans rate a vegan diet as less healthy is because they do not want to admit there is a reason they should follow it. This could be a form of reactance, which is an emotional and cognitive resistance to a message characterized by perceived threat to freedom, anger toward the message, and counterarguments against the message, such as denial (Hall et al., 2017). In other words, non-vegans may feel that their autonomy to choose the foods that they enjoy is being threatened due to all of the positive health messages associated with veganism. Subsequently, in order to avoid this, non-vegans elicit a negative reaction by claiming to believe that the vegan diet is not as healthy as it is made out to be. This is reflected in survey responses from non-vegans such as, "[To be vegan means] to be misguided about nutrition," "They look unhealthy and skinny," or "They look anemic, thin with very little muscle, circles under eyes." These attitudes towards veganism reduce the effectiveness of the messaging around the health benefits of veganism and provide non-vegans with an excuse to continue their lifestyle as is.

Another interesting result from the survey was that vegans rated the vegan diet as more difficult than non-vegans did. This would indicate that even though vegans find their diet difficult, they have motivating factors which keep them dedicated. This could also be an example of the influence strategy of commitment, because it is known that when individuals make a commitment, especially a difficult commitment, they are more likely to follow through with their promise to ensure that the commitment is maintained (Martin & Warner, 2015). In other words, since vegans have committed to following a vegan diet, they are more likely to persevere even though it is difficult. Additionally, previous studies have found that individuals maintain their commitment not only for themselves, but also because they care how others might perceive them if they do not maintain their promise (Martin & Warner, 2015). Therefore, it may be true that vegans feel their lifestyle is difficult but cannot go back in fear of how others would react. Conversely, however, it may be true that vegans want others to know their commitment is difficult so that they will be perceived as superior and having significant willpower for doing it anyway. However, this also raises the question: if non-vegans do not believe it is difficult to follow a vegan diet, then why aren't more doing so? It is possible that non-vegans simply do not care about the concept enough to go vegan, or do not think that any of the benefits are beneficial enough to them. Additionally, those considering veganism rated the diet as most difficult compared to the other two groups. It is possible that this perception of difficulty is what is keeping those considering veganism from actually being vegan.

  The survey also revealed that there were differences in how each group perceived media influence. It seems as though vegans believed media relating to veganism to be positively influential, while non-vegans did not. According to the results from Twitter, the topic of veganism scored "happier" overall compared to all other tweets. This is contrary to what was hypothesized, as it was initially thought that Twitter would portray a more negative picture of veganism. These differences could be due to the fact that vegans are the primary community which are actively engaging in topics surrounding veganism on social media. Therefore, they would be more likely to talk about the subject in a positive manner. Non-vegans, on the other hand, may just not be paying attention to vegan sources of media, or contributing to those sources. It is also possible that non-vegans only receive information about veganism from mainstream sources, which may not be as positive as social media. Future research could attempt to examine vegan portrayals in mainstream sources of media to compare to our findings for vegan social media.

Furthermore, differences in how masculinity is viewed in vegans are apparent between groups. Those who are non-vegan seem to strongly believe that vegans are less masculine than non-vegans, while this belief is less evident in the vegan community. Masculinity related to veganism is a common bias, and this may be an additional reason why non-vegans are not interested in becoming vegan (Thomas, 2016). This trend seemed to be supported on Instagram as well, as about sixty-two percent of the posts containing people were exclusively female. This suggests that the female vegan community is more prominent than the male vegan community on this platform, or that veganism is geared more towards females, which may partly explain where the bias of vegans not being masculine comes from.

As revealed in the coded Instagram posts, the majority of the pictures pertaining to food were everyday foods. This is surprising, as it would seem as though "food porn" type pictures would have made veganism appear more appealing. This could explain why non-vegans are not very influenced by social media, as shown from the survey results. If the food that is being promoted associated with a vegan diet is not appetizing, then it most likely would not convince non-vegans to become vegan. The large amount of everyday food could also indicate that social media related to veganism is mainly aimed at the vegan community. If someone is already vegan, they do not need to be lured into the lifestyle by seeing perfect pictures of food. Instead, they may be more interested in finding ways to maintain their lifestyle in a convenient way. This could explain why so many individuals are sharing pictures of their everyday foods, to inspire other vegans to easily incorporate vegan food into their busy day. If social media surrounding veganism is aimed at the vegan community, this could suggest that non-vegans do not even see much vegan content. Therefore, they are definitely not being influenced.

Additional research is needed to determine what was occurring on the outlier dates from Twitter. It is apparent that World Vegan Day tends to be a "happier" scored day, as individuals are most likely discussing veganism in a positive, celebratory manner. However, specific events that occurred on the happiest and least happy scored days do not appear to coincide with any notable events. It is possible that there were no noteworthy events occurring on these days, and the happiness scores were the result of insignificant retweets. Conducting our survey on MTurk may have limited the population who took our survey or the time they spent on the survey. However, our sample was relatively diverse on MTurk, most likely more diverse than would have been available if we had attempted to recruit our survey participants in-person in our northeast, rural location. However, it may be that those who agree to participate in MTurk surveys are fundamentally different than other populations. Strengths of our research include a multi-method examination of vegan perceptions, as well as survey data from both vegans, non-vegans, and those considering veganism, which allowed for interesting comparisons around vegan perceptions.

**Conclusion**

As previously mentioned, the substantial research revealing the health benefits of the vegan diet is incommensurate with the total number of people that choose to go vegan. A variety of factors account for this discrepancy, and further investigation is needed to estimate their relative weights. However, one possible explanation could be how the subject is portrayed on social media. Veganism is portrayed in a positive manner aimed at vegans, however, it is possible that non-vegans are not exposed to this information or do not want to be exposed. Dietary choices are driven primarily by taste. If non-vegans enjoy the foods they eat, they may have no desire to switch to veganism despite the media attention and associated health benefits. In fact, non-vegans may experience reactance to the idea of becoming vegan, convincing themselves that the vegan diet is unhealthy. If this is the case, instead of convincing people with data and research, it may be more effective to show people that vegan food can taste good too.